# A Rapid GeoSAM-Based Workflow for Multi-Temporal Glacier Delineation: Case Study from Svalbard

*Technical Report*

Alexandru Hegyi

Department of Geosciences, University of Oslo, alexandru.hegyi@geo.uio.no

**Note.** This document is a technical report not a scientific work. It presents a practical workflow for rapid glacier delineation and is intended as a methodological contribution. If the approach proves useful, the report may be cited accordingly.

**Abstract**

*Consistent glacier boundary delineation is essential for monitoring glacier change, yet many existing approaches are difficult to scale across long time series and heterogeneous environments. In this report, we present a GeoSAM-based, semi-automatic workflow for rapid glacier delineation from Sentinel-2 surface reflectance imagery. The method combines late-summer image compositing, spectral-index-based identification of candidate ice areas, prompt-guided segmentation using GeoSAM, and physically based post-processing to derive annual glacier outlines. The workflow is demonstrated in the Ny-Ålesund and Kongsfjorden region of western Svalbard across multiple years of the Sentinel-2 era. Results show that the approach produces spatially coherent and temporally consistent outlines for major glacier bodies, while most errors are associated with small features affected by water bodies, terrain shadows, or high surface variability. The reliance on derived RGB imagery makes the method flexible and transferable to other optical datasets, with improved performance expected at higher spatial resolution. Although user inspection remains necessary to filter incorrect polygons and adjust thresholds for local conditions, the workflow provides a fast and practical alternative for multi-temporal glacier mapping and ice-loss assessment.*

## 1. Introduction

Glaciers are among the most visible and sensitive indicators of climate variability and change (Nesje et al., 2005; Hock and Huss, 2021). Their retreat and thinning influence regional hydrology, ecosystem functioning, and geohazards, and contribute to global sea-level rise (Huss and Hock, 2015). Because glacier change can be rapid and spatially heterogeneous, sustained monitoring of glacier extent remains a core requirement in cryosphere research and climate-impact assessment (Kääb, 2005).. In practice, glacier area and terminus position are often the most accessible metrics available over long time periods, particularly from satellite remote sensing, and they form the basis for glacier inventories, regional assessments, and model evaluation (Raup et al., 2007).

Glacier outlines are not merely cartographic products; they define the spatial domain for many downstream analyses, including estimates of area change, hypsometry, surface elevation change, and mass-balance modeling (Bloch 2007). Long-term monitoring therefore relies on the ability to extract glacier boundaries consistently across time, sensors, and environmental conditions (Kääb, 2005). Even when the physical signal of retreat is strong, inconsistencies in mapping practices, such as differences in seasonal image selection, illumination conditions, or interpretation rules, can introduce spurious change signals that obscure real trends. This issue is particularly critical when the objective is not to produce a single outline for a given date, but to construct a time series of comparable glacier outlines suitable for quantitative change detection (Kääb, 2005).

Despite decades of remote sensing applications, accurate glacier delineation remains challenging in several common scenarios. Seasonal snow can mask glacier margins and create ambiguous transitions between transient snow cover and perennial ice (Paul et al., 2013). Clouds, cirrus, and cloud shadows can generate false edges that mimic glacier boundaries, while topographic shading and varying solar geometry can reduce contrast along valley walls and accumulation areas (Paul et al., 2013; Ye et al., 2024). Debris-covered ice presents a distinct challenge, as its spectral signature can approach that of surrounding moraines and bedrock, making simple spectral separation unreliable. Tidewater glaciers introduce additional complexity, as calving fronts may change rapidly and be influenced by short-term dynamic processes, complicating the extraction of temporally consistent boundaries (Chen et al., 2023).

Historically, manual digitization has been treated as the reference approach for glacier mapping, as expert interpretation can often resolve visually ambiguous boundaries. However, manual delineation is time-consuming, difficult to reproduce at scale, and inherently subjective, particularly in areas affected by snow cover, debris, or low contrast. These limitations become increasingly restrictive when attempting to generate multi-year or multi-decadal datasets, or when extending monitoring consistently across larger regions (Paul et al., 2013; Ye et al., 2024).

A wide range of semi-automated glacier mapping approaches have therefore been proposed. Spectral-index methods, such as band ratios and normalized difference indices, can be effective for clean ice and snow under favorable conditions and remain widely used due to their simplicity and interpretability (Paul et al., 2013). Object-based image analysis (OBIA) methods incorporate spatial context and can reduce pixel-level noise but often depend on parameter choices that do not transfer robustly across years or illumination conditions (Zhang et al., 2019). More recently, convolutional neural networks (CNNs) and related deep learning approaches have shown promising performance where extensive labeled training data are available and well matched to the target region and sensor characteristics (Baumhoer et al., 2019).

However, both classical and CNN-based methods face practical scaling barriers for long time series and heterogeneous geographic settings. Threshold-based approaches typically require returning across seasons, sensors, and snow conditions, while CNN-based models depend on large, consistently labeled datasets and may generalize poorly outside their training domain. These constraints are particularly limited for long-term glacier monitoring, where training data may be sparse, inconsistent across time, or unavailable for many regions of interest.

Foundation models for image segmentation offer an alternative pathway (Özdemir et al., 2024). Rather than training glacier-specific models from scratch, foundation models can be prompted to segment target objects in a largely zero-shot or low-shot manner (Moghimi et al., 2024). The Segment Anything Model (SAM) and related approaches have demonstrated broad generalization across diverse image content, enabling segmentation guided by simple prompts such as points or bounding boxes (Kirillov et al., 2023). For glacier mapping, this paradigm is appealing because glacier ice and snow exhibit visually coherent characteristics across many settings, and because long time series benefit from methods that do not require repeated retraining as sensors or conditions change.

Nevertheless, the application of foundation segmentation models to systematic glacier monitoring is not trivial. Prompt selection can introduce a new source of subjectivity, and multi-temporal change analysis requires consistent prompting strategies to avoid artificial variability. In addition, geospatial workflows must address practical challenges such as large scene sizes, image tiling, georeferencing, and the conversion of raster masks into vector glacier outlines. GeoSAM, an adaptation of SAM designed for geospatial imagery, provides a practical framework for applying prompt-based segmentation to satellite data, yet its suitability for glacier boundary delineation and for producing temporally consistent outline time series has not been systematically evaluated in a glacier-focused context (Ibn Sultan et al., 2023).

This report presents a GeoSAM-based workflow for semi-automatic glacier delineation from Sentinel-2 imagery and evaluates its potential as a data-enabling tool for multi-temporal glacier monitoring. Rather than aiming to produce a complete regional glacier inventory, the focus is methodological: to assess whether a prompt-driven foundation model approach can generate reliable and repeatable glacier outlines across multiple years, and to identify strengths, limitations, and safeguards necessary for scalable application.

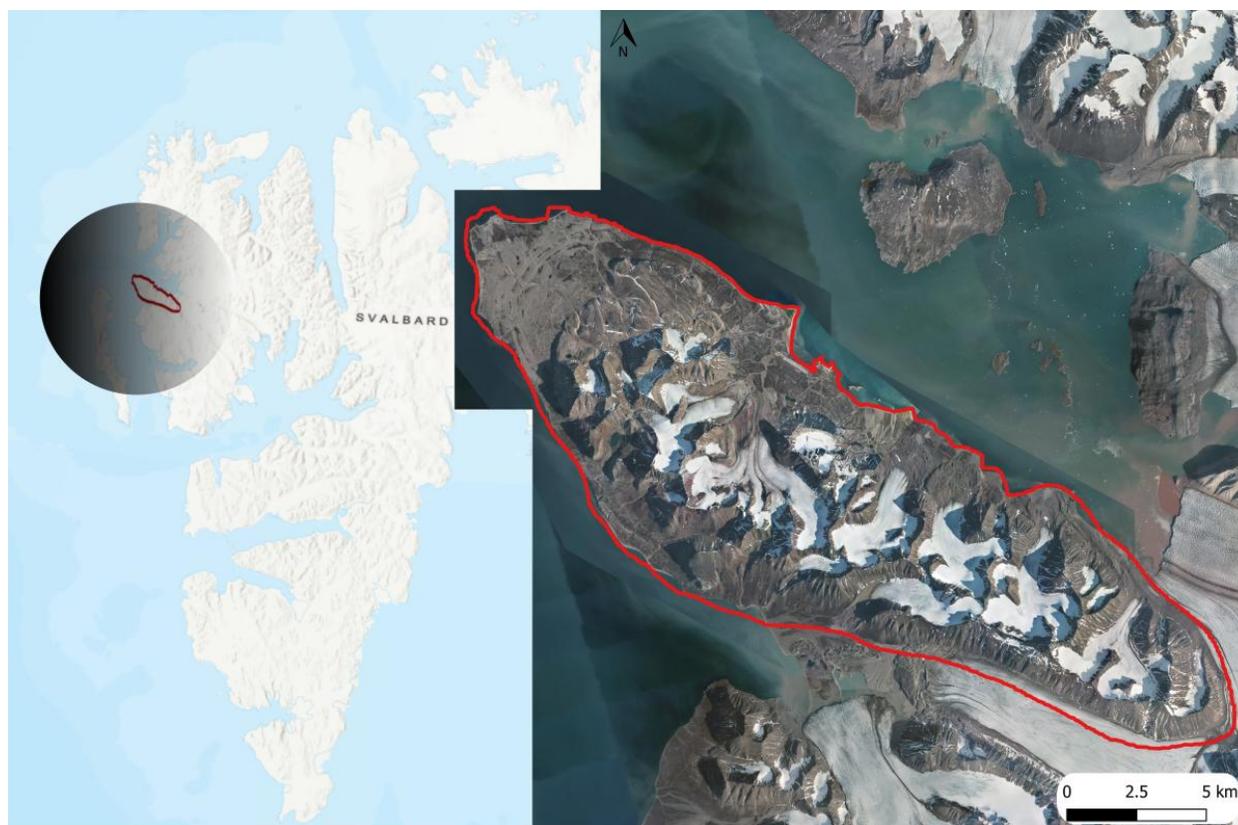

**Figure 1.** Location of the study area in western Svalbard and overview of the analyzed region. The inset map shows the position of the study area within the Svalbard archipelago, while the main panel displays a Sentinel-2 true-color composite of the Ny-Ålesund–Kongsfjorden region. The red outline delineates the area of interest used for glacier mapping and analysis.

We demonstrate the proposed workflow in the Ny-Ålesund–Kongsfjorden region of western Svalbard (Figure 1), a well-studied Arctic test area characterized by a high density of glaciers, diverse glacier geometries, and extensive observational records. Rather than focusing on a limited set of representative glaciers, the analysis is conducted at the regional scale, encompassing land-terminating valley glaciers, polythermal systems, and marine-terminating tidewater glaciers with varying degrees of dynamical complexity. This regional approach allows assessment of segmentation robustness across a wide range of boundary conditions, surface characteristics, and temporal behaviours.

Using multi-temporal Sentinel-2 surface reflectance imagery acquired during consistent late-summer windows, we (1) implement a repeatable preprocessing and compositing strategy optimised for glacier mapping, (2) design an automated prompt-generation scheme suitable for large-scale time-series segmentation, (3) apply the GeoSAM framework with geospatial tiling and physically informed post-processing to derive georeferenced glacier outlines, and (4) introduce automated quality-control indicators to identify years and locations where segmentation results are likely affected by clouds, seasonal snow, noise, or abrupt area changes.

The specific objectives of this study are to:
- develop an end-to-end GeoSAM-based workflow that converts multi-temporal Sentinel-2 imagery into glacier masks and vector outlines suitable for regional change analysis.
- evaluate segmentation robustness and temporal consistency across heterogeneous glacier types within a complex Arctic fjord system.

By framing GeoSAM as a practical, quality-controlled segmentation component within a geospatial monitoring pipeline, this study provides a foundation for scaling from regional Arctic test sites to broader regional and, ultimately, global glacier time-series products.

## 2. Methods

This report employs the Segment Anything Model (SAM) as a general-purpose (Figure 2), prompt-based image segmentation model and integrates it within a geoscience-oriented framework referred to as GeoSAM (Ibn Sultan et al., 2023). SAM is a foundation model designed for object-agnostic segmentation, in which spatially coherent image regions are delineated based on sparse prompts (e.g. points or bounding boxes) rather than predefined semantic classes. As a purely vision-based model, SAM is used here exclusively for geometric boundary refinement (Kirillov et al., 2023)..

GeoSAM represents the adaptation of SAM to geospatial remote sensing applications by embedding the model within a physically informed, reproducible workflow (Ibn Sultan et al., 2023). In GeoSAM, segmentation is guided by expert-defined prompts derived from spectral indices and spatial priors, and SAM outputs are subsequently constrained using physically interpretable criteria (Ibn Sultan et al., 2023). This separation between visual boundary extraction and physical feature identification ensures spatial consistency while maintaining scientific interpretability.

Annual glacier extents were derived from Sentinel-2 Level-2A surface reflectance imagery. To minimize seasonal snow contamination and maximize spectral contrast between glacier ice and surrounding terrain, we restricted the analysis to a late-summer temporal window (10–31 August). All scenes intersecting the study area were filtered for clouds and cirrus using the QA60 quality band.

For each year, a 25th-percentile (P25) composite was computed from all cloud-filtered observations within the temporal window. The P25 statistic effectively suppresses residual cloud, haze, and shadow artefacts while preserving the radiometric properties of glacier ice. The resulting annual composite included six spectral bands: B2 (blue), B3 (green), B4 (red), B8 (near-infrared), B11 and B12 (short-wave infrared). All raster products were projected to EPSG:25833 (UTM Zone 33N) at 10 m spatial resolution.

From each annual P25 composite, we computed the Normalized Difference Snow Index (NDSI) and the Normalized Difference Vegetation Index (NDVI) using standard formulations. These indices were used to generate an initial binary ice mask through conservative thresholding, followed by morphological filtering and minimum-area constraints to remove small, isolated patches.

The ice mask represents candidate snow and ice surfaces and is not interpreted as a final glacier delineation. Instead, it serves as (i) a spatial prior for segmentation, (ii) a basis for prompt generation, and (iii) a reference set for adaptive threshold estimation.

All annual raster products (P25 composite, NDSI, NDVI, ice-mask) were organized into a standardized directory structure with one folder per year. This ensured consistent access patterns and reproducibility across all subsequent processing steps.

Candidate glacier regions were identified by extracting connected components from the ice-mask. Components below a minimum area threshold (5,000 m²) were excluded. For each remaining component, we generated: a bounding box expanded by a fixed spatial padding to provide contextual information, multiple positive point prompts sampled within the component, negative point prompts sampled in a surrounding buffer zone.

Negative prompts were specifically designed to suppress segmentation leakage into adjacent non-glacier surfaces, such as water bodies, bedrock, and shadowed terrain. Bounding boxes and prompt points were stored as vector datasets and used as explicit inputs to the segmentation model.

To enable segmentation with a vision-based model, each annual P25 composite was converted into a true-color RGB image using Sentinel-2 bands B4 (red), B3 (green), and B2 (blue). Surface reflectance values were linearly rescaled to 8-bit radiometric depth (0–255) using a fixed reflectance range, preserving relative contrast while ensuring compatibility with the segmentation framework.

Glacier boundary refinement was performed using the Segment Anything Model (SAM) with a Vision Transformer backbone. For each year, the RGB image was loaded once into the predictor. Segmentation was then performed independently for each candidate glacier region using the corresponding bounding box and point prompts.

A single segmentation mask was produced per candidate region. In this workflow, SAM is employed exclusively for geometric boundary refinement and is not used to infer glacier class membership.

Raw SAM predictions were filtered using physically based spectral constraints to remove false positives associated with water surfaces, terrain shadows, and low-albedo non-ice features.

Thresholds for NDSI, NDVI, and short-wave infrared reflectance (B11) were automatically derived on a per-year basis from pixels classified as ice in the ice-mask. Specifically: a lower-percentile NDSI value defined the minimum snow/ice threshold, an upper-percentile NDVI value defined the maximum vegetation threshold, an upper-percentile B11 value defined the maximum allowable SWIR reflectance for ice.

Small safety margins were applied to prevent over-constraining, and thresholds were restricted to physically plausible ranges. This adaptive strategy accounts for interannual variability in illumination conditions, surface melt, and atmospheric effects.

A pixel was retained as glacier ice only if it satisfied all the following conditions: classified as foreground by SAM, NDSI above the auto-derived threshold, NDVI below the auto-derived threshold, B11 reflectance below the auto-derived threshold, valid RGB radiometric values, optional inclusion within the ice-mask.

Filtered segmentation masks from all candidate regions were merged into a single annual glacier mask. Morphological operations were applied to remove small, isolated objects and to fill small holes, improving spatial coherence while preserving glacier extent.

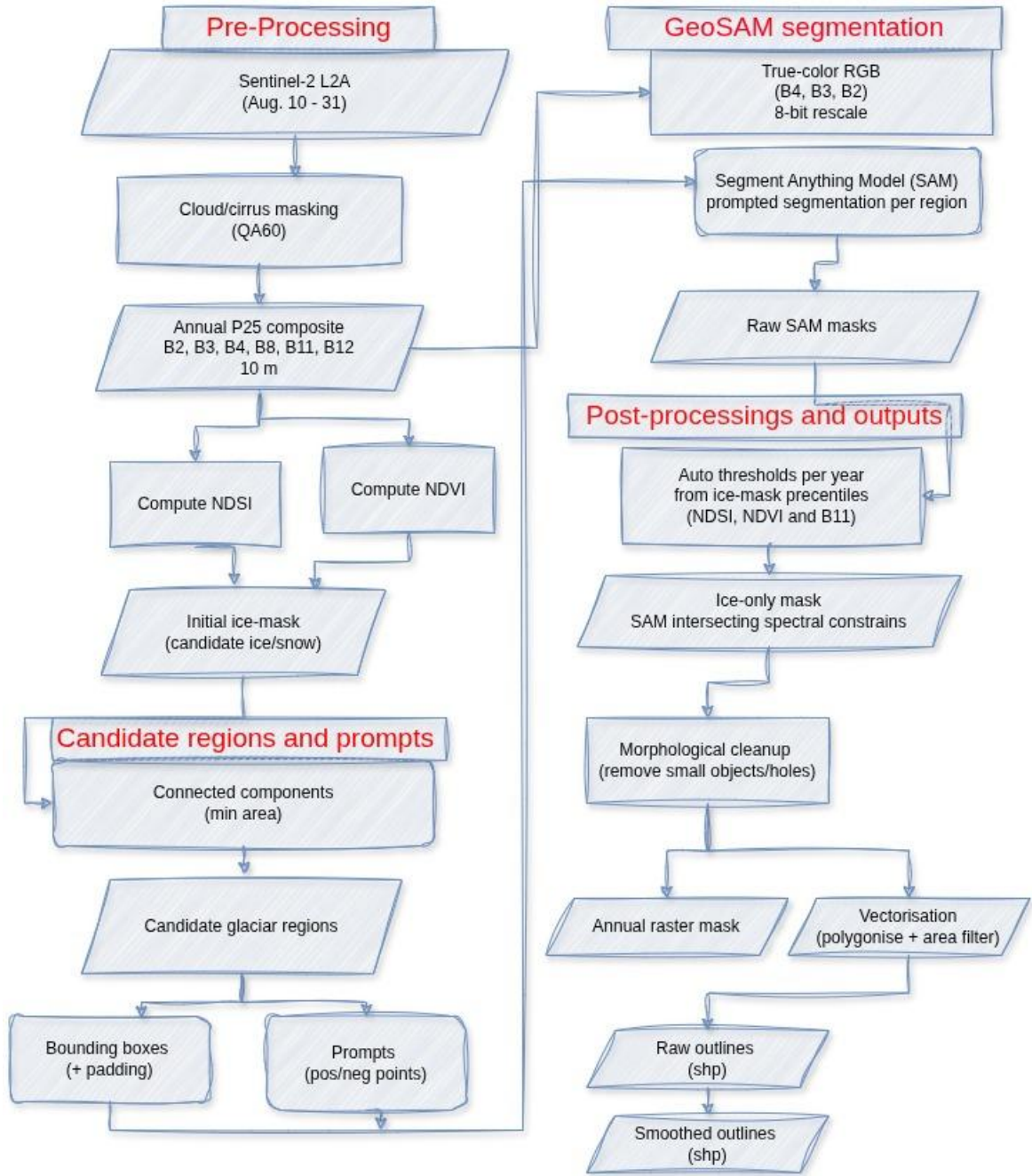

**Figure 2.** Schematic overview of the GeoSAM-based glacier delineation workflow. The process consists of three main stages: pre-processing of Sentinel-2 imagery, identification of candidate glacier regions and generation of segmentation prompts, and GeoSAM-based boundary refinement followed by post-processing. Spectral indices are used to define candidate ice areas and to derive physically based constraints, while the Segment Anything Model (SAM) is applied exclusively for geometric segmentation. Final outputs include annual raster glacier masks and both raw and smoothed vector glacier outlines.

Annual glacier masks were converted to vector polygons. Polygons below a minimum area threshold were removed. To reduce pixel-stair artefacts inherent in raster-derived outlines, polygon boundaries were smoothed using a morphological buffer–unbuffed operation, followed by optional topology-preserving simplification.

Smoothing parameters were held constant across all years to ensure temporal consistency in glacier area estimates. Both raw and smooth outlines were retained and exported for analysis.

For each year, the workflow produces: a raster glacier mask, raw glacier outline polygons, smoothed glacier outline polygons.

These products enable robust interannual glacier area analysis and integration with GIS-based cryosphere studies.

The proposed approach (Figure 2) combines deep-learning-based boundary refinement with physically interpretable spectral constraints, explicitly separating geometric segmentation from glacier identification. This hybrid strategy improves robustness in shadow-affected and coastal environments and provides a scalable, reproducible framework for long-term glacier monitoring.

## 3. Results

Figure 3 presents a multi-year visual overview of the Sentinel-2 P25 composites together with grayscale representations of spectral indices derived for the study area. The true-color composites illustrate the spatial relationship between glacier ice, exposed bedrock, and surrounding terrain under varying illumination and surface conditions. The corresponding NDSI panels highlight snow and ice surfaces as consistently high-contrast features, clearly separating glacierized areas from non-snow surfaces even in shadow-affected regions. In contrast, NDVI values remain low over glacier ice and increase over vegetated or debris-covered terrain, supporting the discrimination between ice and non-ice surfaces. While interannual variations in reflectance and shadow extent are visible, the combined behavior of the RGB imagery, NDSI, and NDVI remains spatially coherent across all years. This consistency provides a reliable spectral basis for the subsequent glacier delineation and supports the stability of the resulting glacier outlines.

A qualitative accuracy assessment based on visual inspection indicates that most incorrectly delineated polygons are associated with water bodies and terrain shadows (Figure 4). In 2018, 12 out of 43 polygons (≈28%) were identified as incorrect, while in 2020, 17 out of 44 polygons (≈39%) were misclassified. For 2022, 11 out of 49 polygons (≈22%) were incorrect, and in 2024, 13 out of 44 polygons (≈30%) showed evident delineation errors. Most errors correspond to small, spatially isolated features, whereas the main glacier bodies are consistently and accurately delineated across all analyzed years.

Visual inspection of the mapped outlines (Figure 4) shows that misclassifications predominantly occur in areas of open water and in shadowed terrain adjacent to glacier margins. These surfaces exhibit low reflectance and spectral behavior that partially overlaps with glacier ice, particularly under low illumination conditions. Years characterized by higher radiometric variability and stronger shadow contrasts, such as 2022, show an increased sensitivity to these effects, resulting in additional false positives despite the applied spectral constraints.

The results further indicate that the relative impact of these errors decreases when glacier extent is analyzed at broader spatial scales. While small-scale misclassifications affect individual polygons, the overall spatial pattern and extent of major glacier systems remain stable and physically plausible. Consequently, the workflow is well suited for regional-scale glacier mapping and interannual comparison, while caution is required when interpreting small or marginal features in shadow- or water-influenced environments.

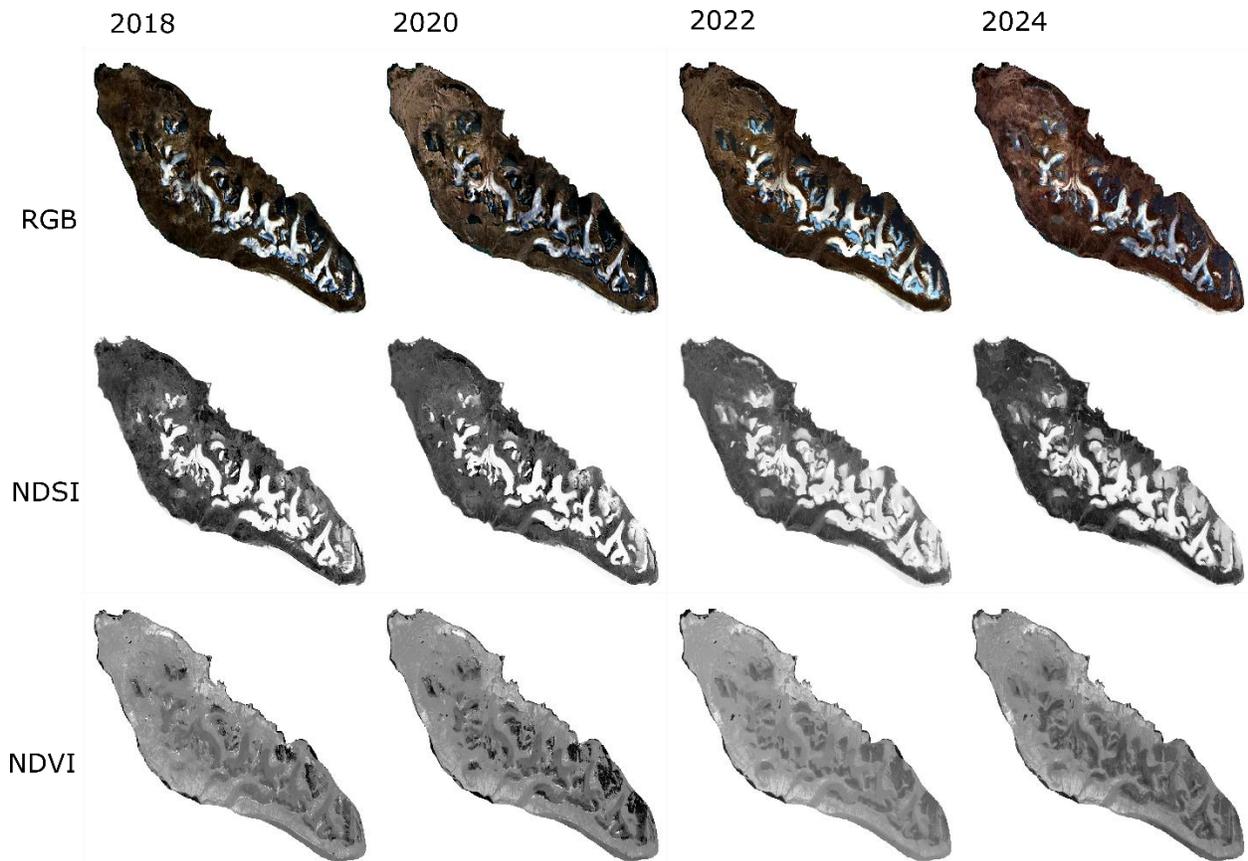

**Figure 3.** Multi-year overview of Sentinel-2 P25 composites and derived spectral indices for the study area. The top row shows true-color composites (B4–B3–B2), while the lower rows illustrate grayscale representations of the Normalized Difference Snow Index (NDSI) and the Normalized Difference Vegetation Index (NDVI) for the analyzed years. Glacier ice remains consistently distinguishable from surrounding terrain despite interannual variations in illumination and surface conditions, providing a stable spectral basis for glacier delineation.

A comparison between glacier outlines derived for 2018 and 2024 (Figure 5) demonstrates that, when observations are sufficiently separated in time and surface conditions are relatively stable, the GeoSAM-based segmentation performs robustly. In these cases, glacier margins are well defined, and the extracted outlines closely follow visible ice boundaries with limited interference from shadows or water-adjacent areas. The reduced short-term radiometric variability between temporally distant observations improves boundary contrast, allowing the segmentation model to better distinguish persistent glacier ice from transient surface features. As a result, the workflow is well suited for detecting long-term glacier area changes and provides a reliable basis for assessing ice loss over time at regional scales.

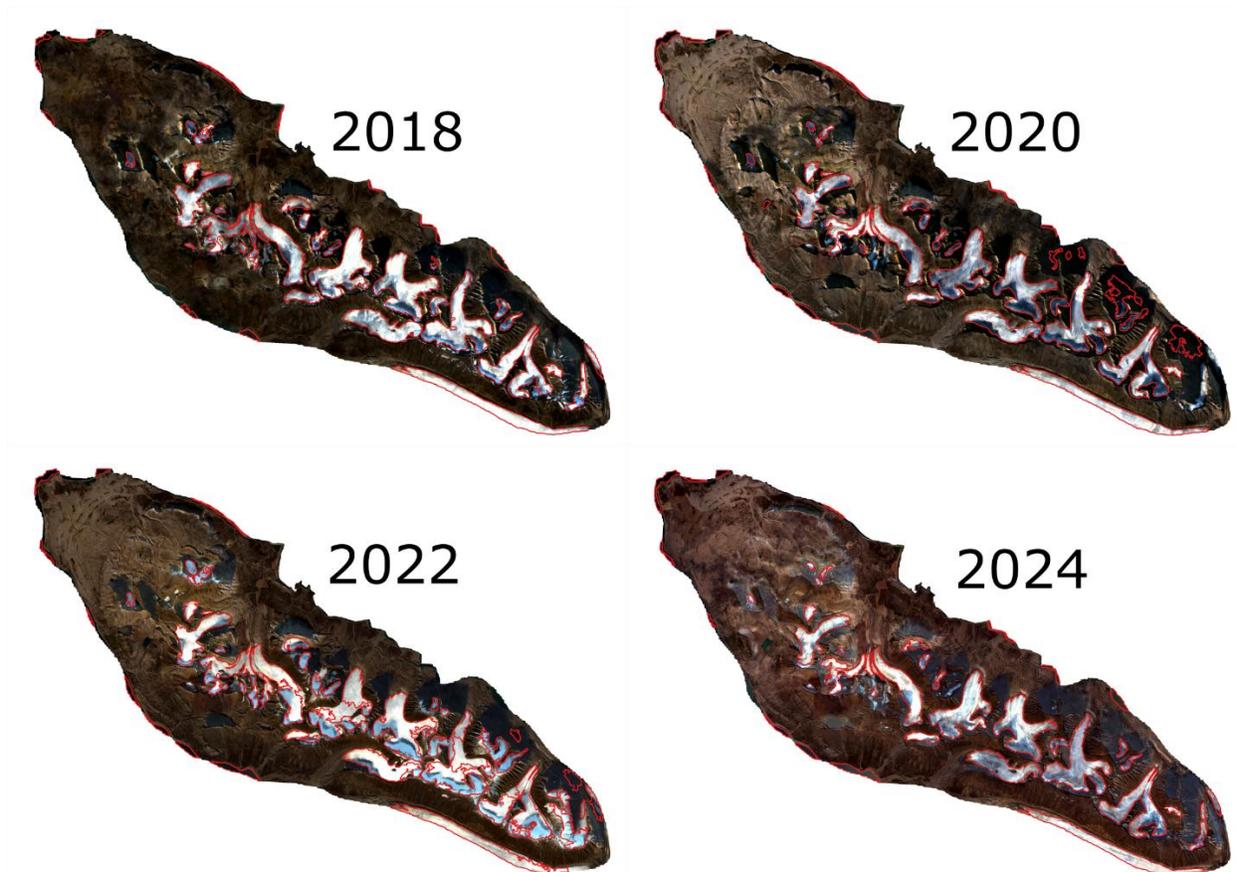

**Figure 4.** Glacier outlines derived for selected years (2018, 2020, 2022, and 2024) overlaid on Sentinel-2 true-color composites. Correctly delineated glacier ice is shown alongside areas of misclassification, which predominantly occur in water-adjacent zones and shadowed terrain. Increased surface and illumination variability in certain years contributes to additional segmentation errors, particularly for small, isolated features.

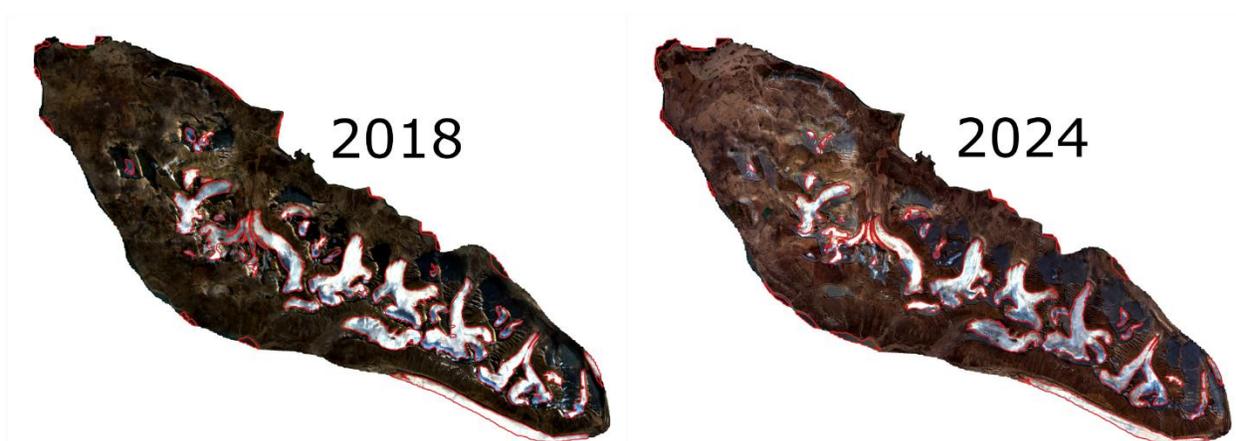

**Figure 5.** Glacier outlines derived for 2018 and 2024 overlaid on Sentinel-2 true-color composites. Red polygons indicate the smoothed glacier outlines produced by the GeoSAM-based workflow. When images are separated by a longer time interval and exhibit relatively low short-term surface variability, glacier boundaries remain clearly defined and spatially consistent, enabling reliable identification of glacier extent changes.

## 4. Brief discussions

The presented methodology demonstrates that a GeoSAM-based workflow provides a fast and flexible alternative for glacier delineation from optical satellite imagery. By combining prompt-guided segmentation with physically interpretable spectral constraints, the approach enables rapid extraction of glacier outlines without the need for extensive model training or complex preprocessing chains. This makes the method particularly suitable for exploratory analyses, regional assessments, and applications where computational efficiency and adaptability are priorities.

An important advantage of the workflow is its reliance on RGB imagery derived rather than sensor-specific spectral inputs. As the segmentation model operates on true-color images, the approach can be applied to any optical dataset for which a representative RGB composite can be generated. This sensor-agnostic design increases the potential applicability of the method across different satellite platforms and time periods. Furthermore, improvements in spatial resolution are expected to directly translate into more accurate glacier boundaries, particularly along complex margins and narrow glacier tongues where mixed pixels currently limit delineation precision.

The results indicate that the method performs best under conditions of relatively low surface variability and clear radiometric contrast between ice and surrounding terrain. When such conditions are met, especially in temporally well-separated observations, the extracted outlines are spatially consistent and well suited for analyzing long-term glacier change. However, residual errors associated with water bodies, terrain shadows, and low-contrast surfaces remain, particularly for small and isolated features. As a result, some level of user interaction or post-processing inspection is currently required to identify and filter incorrect polygons.

Although the workflow is conceptually applicable at global scales, its performance depends on the quality of the input imagery and the appropriateness of the applied spectral and physical thresholds. Adjusting these thresholds to account for regional differences in illumination, surface conditions, debris cover, or melt state would likely improve glacier detectability in specific environments. Consequently, while the method offers a scalable and efficient framework for glacier delineation, optimal results are achieved when it is combined with expert oversight and regionally informed parameter tuning.

Overall, the proposed approach represents a practical compromise between automation and interpretability. It enables rapid glacier mapping across large areas while retaining sufficient flexibility to adapt to varying environmental conditions, making it a valuable tool for regional glacier monitoring and preliminary assessments of ice loss.

## 5. Conclusions

This report demonstrates that a GeoSAM-based, semi-automatic workflow provides a fast and flexible alternative for glacier boundary delineation from optical satellite imagery. By combining prompt-guided segmentation with physically interpretable spectral constraints, the method produces spatially coherent and temporally consistent glacier outlines for major glacier bodies while limiting false detections in most cases.

The results show that the approach is particularly effective when applied to temporally well-separated observations with relatively stable surface conditions, enabling reliable identification of long-term glacier area changes. Although residual errors remain in areas affected by water bodies, terrain shadows, and high surface variability, these are largely confined to small, isolated features and can be addressed through limited user inspection.

Overall, the workflow offers a practical compromise between automation and interpretability. Its reliance on derived RGB imagery makes it adaptable to different optical sensors, and improvements in spatial resolution are expected to further enhance boundary accuracy. The method is therefore well suited for regional glacier monitoring and exploratory assessments of glacier change, while continued refinement of spectral thresholds and user-guided filtering will be important for broader-scale applications.

## 6. References

Baumhoer, C. A., Dietz, A., Kneisel, C., & Kuenzer, C. (2019). Automated extraction of Antarctic glacier and ice shelf fronts from Sentinel-1 imagery using deep learning. *Remote Sensing, 11*(21), Article 2529. https://doi.org/10.3390/rs11212529

Bolch, T. (2007). Climate change and glacier retreat in northern Tien Shan (Kazakhstan/Kyrgyzstan) using remote sensing data. *Global and Planetary Change, 56*(1–2), 1–12. https://doi.org/10.1016/j.gloplacha.2006.07.009

Chen, J., Gao, H., Han, L., Yu, R., & Mei, G. (2023). Susceptibility analysis of glacier debris flow based on remote sensing imagery and deep learning: A case study along the G318 Linzhi section. *Sensors, 23*(14), Article 6608. https://doi.org/10.3390/s23146608

Hock, R., & Huss, M. (2021). Glaciers and climate change. In *Climate change* (pp. 157–176). Elsevier. https://doi.org/10.1016/B978-0-12-821575-3.00009-0

Huss, M., & Hock, R. (2015). A new model for global glacier change and sea-level rise. *Frontiers in Earth Science, 3*, Article 54. https://doi.org/10.3389/feart.2015.00054

Ibn Sultan, R., Li, C., Zhu, H., Khanduri, P., Brocanelli, M., & Zhu, D. (2023). GeoSAM: Fine-tuning SAM with sparse and dense visual prompting for automated segmentation of mobility infrastructure. *https://arxiv.org/html/2311.11319v2*.

Kääb, A. (2005). Combination of SRTM3 and repeat ASTER data for deriving alpine glacier flow velocities in the Bhutan Himalaya. *Remote Sensing of Environment, 94*(4), 463–474. https://doi.org/10.1016/j.rse.2004.11.003

Kirillov, A., Mintun, E., Ravi, N., Mao, H., Rolland, C., Gustafson, L., Xiao, T., Whitehead, S., Berg, A. C., Lo, W.-Y., Dollár, P., & Girshick, R. (2023). Segment anything. In *Proceedings of the IEEE/CVF International Conference on Computer Vision*. https://arxiv.org/abs/2304.02643.

Moghimi, A., Welzel, M., Çelik, T., & Schlurmann, T. (2024). A comparative performance analysis of popular deep learning models and the Segment Anything Model (SAM) for river water segmentation in close-range remote sensing imagery. *IEEE Access, 12*, 52067–52085. https://doi.org/10.1109/ACCESS.2024.3385425

Nesje, A., & Dahl, S. O. (2005). Glaciers as indicators of Holocene climate change. In *Global change in the Holocene*. Routledge.

Özdemir, S., Akbulut, Z., Karslı, F., & Kavzoğlu, T. (2024). Extraction of water bodies from high-resolution aerial and satellite images using visual foundation models. *Sustainability, 16*(7), Article 2995. https://doi.org/10.3390/su16072995

Paul, F., Barrand, N. E., Baumann, S., Berthier, É., Bolch, T., Casey, K. A., et al. (2013). On the accuracy of glacier outlines derived from remote-sensing data. *Annals of Glaciology, 54*(63), 171–182. https://doi.org/10.3189/2013AoG63A296

Raup, B., Raciviteanu, A., Khalsa, S. S., Helm, C., Armstrong, R. L., & Arnaud, Y. (2007). The GLIMS geospatial glacier database: A new tool for studying glacier change. *Global and Planetary Change, 56*(1–2), 101–110. https://doi.org/10.1016/j.gloplacha.2006.07.018

Ye, Q., Wang, Y., Liu, L., Guo, L., Zhang, X., Dai, L., et al. (2024). Remote sensing and modeling of the cryosphere in High Mountain Asia: A multidisciplinary review. *Remote Sensing, 16*(10), Article 1709. https://doi.org/10.3390/rs16101709

Zhang, E., Liu, L., & Huang, L. (2019). Automatically delineating the calving front of Jakobshavn Isbræ from multitemporal TerraSAR-X images: A deep learning approach. *The Cryosphere, 13*(6), 1729–1741. https://doi.org/10.5194/tc-13-1729-2019

## 7. Appendix A: Data and code availability

All scripts and example data used in this report are available in a public GitHub repository:

**https://github.com/alexandruhegyi/A-Rapid-GeoSAM-Based-Workflow-for-Multi-Temporal-Glacier-Delineation**

The repository contains the full workflow required to reproduce the glacier delineation results, including preprocessing of Sentinel-2 imagery, prompt generation, GeoSAM-based segmentation, and post-processing of glacier outlines. Input data are organized by year, and scripts are provided to run the workflow sequentially.

To apply the method to a new study area, users are required to (i) provide Sentinel-2 surface reflectance imagery for a consistent late-summer period, (ii) generate annual RGB composites and spectral indices, and (iii) adjust basic spectral thresholds where necessary to account for local conditions. The workflow is designed for reproducibility and can be adapted to other optical datasets with minimal modification.